\documentclass[osajnl,twocolumn,showpacs,superscriptaddress,10pt]{revtex4-1} %% use 11pt for Applied Optics
\usepackage{amsmath,amssymb,graphicx}
\usepackage{braket}
\newcommand{\sgn}{\mathop{\mathrm{sgn}}}
\usepackage{bm}% bold math
\usepackage{multibib}
\usepackage{filecontents}
\newcites{ref}{References}
\usepackage{color}

\usepackage{soul}
\usepackage{ulem}
\usepackage{siunitx}
\usepackage{lineno}
\makeatletter
\renewcommand\@biblabel[1]{#1.}
\makeatother

\makeatletter
\def\fixedlabel#1#2{%
  \@bsphack
  \begingroup
    \@onelevel@sanitize\@currentlabelname
    \edef\@currentlabelname{%
      \expandafter\strip@period\@currentlabelname\relax.\relax\@@@%
    }%
    \phantomsection%
    \protected@write\@auxout{}{%
      \string\newlabel{#1}{%
        {#2}%
        {\thepage}%
        {#2}%
        {\@currentHref}{}%
      }%
    }%
  \endgroup
  \@esphack
}
\makeatletter

\begin{document}

\title{Transformation Optics scheme for two-dimensional materials }

\author{Anshuman Kumar}
\affiliation{Mechanical Engineering Department, Massachusetts Institute of Technology, Cambridge, MA - 02139}
\author{Kin Hung Fung}
\affiliation{Department of Applied Physics, The Hong Kong Polytechnic University, Hong Kong}
\author{M. T. Homer Reid}
\affiliation{Mathematics Department, Massachusetts Institute of Technology, Cambridge, MA - 02139}
\author{Nicholas X. Fang}\email{Corresponding author: nicfang@mit.edu}
\affiliation{Mechanical Engineering Department, Massachusetts Institute of Technology, Cambridge, MA - 02139}

\begin{abstract}Two dimensional optical materials, such as graphene can be characterized by a surface conductivity. So far, the transformation optics schemes have focused on three dimensional properties such as permittivity $\epsilon$ and permeability $\mu$. In this paper, we use a scheme for transforming surface currents to highlight that the surface conductivity transforms in a way different from $\epsilon$ and $\mu$. We use this surface conductivity transformation to demonstrate an example problem of reducing scattering of plasmon mode from sharp protrusions in graphene. 

\end{abstract}

% REPLACE WITH CORRECT OCIS CODES FOR YOUR ARTICLE
                          % NOTE: \ocis{} IS ALIASED TO \pacs{} BUT MUST
                          % FORMAT THE TERMS CORRECTLY FOR EACH JOURNAL

\maketitle %% required

%\section{Introduction}

%\linenumbers
Transformation optics\cite{Pendry03082012, doi:10.1021/nl100800c} has proved to be a powerful technique to control propagation of electromagnetic waves. The basic idea is that under coordinate transformations, Maxwell's equations remain form-invariant, provided the material parameters are appropriately modified. Then the wave electromagnetic wave propagation in the transformed medium can be thought of as occurring in the original medium. Numerous applications of this technique have been proposed and demonstrated. Some examples are invisiblity cloaks\cite{Schurig10112006, Liu16012009, DielectricCloak}, waveguides with sharp bends\cite{roberts:251111}, subwavelength image manipulation\cite{Schurig:07}, etc.

With the recent discovery of two dimensional materials such as graphene\cite{doi:10.1038/nmat1849} and MoS$_2$\cite{MoS2review,  PhysRevLett.105.136805}, there is an enormous interest in studying their optical properties\cite{Hamm14062013}. Such materials are usually characterized by a surface conductivity instead of the volume conductivity which describes the usual three dimensional materials. As such it is expected that the techniques of transformation optics, as is usually employed, will have to be modified to take into account the change in the dimensionality of the material parameter. To our knowledge, this is the first time that a full transformation optics scheme involving surface conductivity has been considered. Although implementing transformation optics using spatial modulation of surface conductivity has been proposed in \cite{GrapheneTransfOptics}, we note that they do not talk about transformations which can take the graphene out of the plane. In other words, they only consider flat graphene.

In this paper, we will show how surface conductivity transforms under arbitrary coordinate transformations. We will find that the transformation rule is indeed different compared to the bulk conductivity. Then we will present an example to show how this surface conductivity transformation works for the simple case of a two dimensional transformation. We show how to reduce plasmon scattering from a triangular protrusion in graphene. It is indeed possible to implement such a surface conductivity transformation in graphene using gate voltage\cite{10.1126/science.1102896} and chemical doping.

%\section{Transformation Optics for Surface Conductivity }

Let us consider the case of a 2D material sitting in between two dielectric materials. The source free Amp{\`e}re's Law is written in this case as:
\begin{equation}
\nabla\times\mathbf{H} = \epsilon_0\hat{\epsilon}_{bulk}\frac{\partial{\mathbf{E}}}{\partial{t}} +  \frac{\partial{\mathbf{P}_s}}{\partial{t}}|\nabla F(\mathbf{r})|\delta(F(\mathbf{r}))
\label{eq: Ampere_with_Surface}
\end{equation}
where $F(\mathbf{r}) = 0$ is the equation of the surface discontinuity, $\epsilon$ is the permittivity of the surrounding three dimensional materials and $\mathbf{P}_s$ is the surface polarization of the two dimensional material. The time derivative of this surface polarization gives rise to a surface current density:
\begin{equation}
\mathbf{J}_s = \frac{\partial{\mathbf{P}_s}}{\partial{t}} = \sigma^{2D}\mathbf{E}_{||} = \hat{\sigma}\mathbf{E}
\label{eq:surface_ohms_law}
\end{equation}
\begin{figure*}[htb]
\centering\includegraphics[width=10cm]{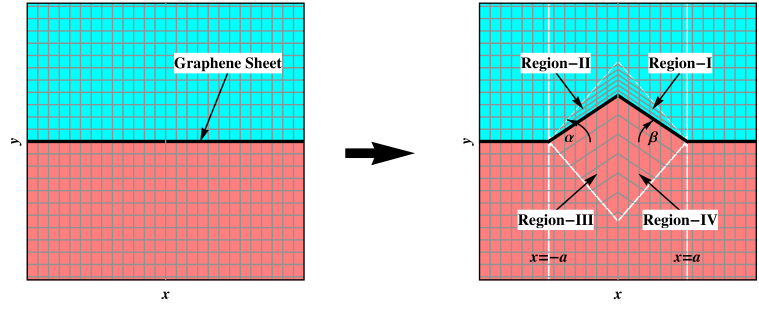}
\caption{Coordinate transformation which compresses a triangular region towards the top and rarefies it at the bottom. A sharp protrusion in the graphene is produced at the center.}
\label{fig:Schematic_Graphene_Transformation}
\end{figure*}
For the usual case of the interface of two bulk materials, this term is absent, however for 2D materials we need to retain this term in the derivation of the boundary conditions due to the presence of the Dirac delta function in Eq.\ref{eq: Ampere_with_Surface}. Thus, in this case of a finite surface electrical conductivity $\sigma^{2D}$, the boundary condition for the tangential magnetic field can be written in terms of a surface current\cite{PhysRevB.80.245435}
\begin{equation}
\hat{\mathbf{n}}\times \Delta\mathbf{H} =\mathbf{J_s} = \hat{\sigma}\mathbf{E}
\label{eq:Surface_Current_BC}
\end{equation}
where $\mathbf{J}_s$ is the surface current density, $\hat{\mathbf{n}}$ is the unit normal to the surface and $\hat{\sigma}$ is the surface conductivity tensor. For instance, an isotropic surface conductivity would be represented in terms of the basis \{$\Ket{t_1}$, $\Ket{t_2}$, $\Ket{n}$\}, as $\hat{\sigma} = \sigma^{2D} (\Ket{t_1}\Bra{t_1} + \Ket{t_2}\Bra{t_2} )$. Here $\Ket{t_1}$ and $\Ket{t_2}$ are the two orthonormal local tangential vectors and $\Ket{n}$ is the local normal unit vector. Our aim here is to find out the transformation rule for the $\hat{\sigma}$ tensor.

Based on Eq.\ref{eq: Ampere_with_Surface} and \ref{eq:surface_ohms_law}, the permittivity can be thought of as containing a Dirac delta function: 
\begin{equation}
\mathbf{\hat{\epsilon}} = \mathbf{\hat{\epsilon}}_{bulk} + \frac{i|\nabla F(\mathbf{r})|\delta(F(\mathbf{r}))}{\omega\epsilon_0} \hat{\sigma}
\label{eq:Generalized_epsilon}
\end{equation}
Now, applying the usual transformation rule for permittivity $\mathbf{\hat{\epsilon}'}=\Lambda\hat{\epsilon}\Lambda^T/\det(\Lambda)$ to Eq.\ref{eq:Generalized_epsilon} and the standard rule for the change of variables in a delta function, we arrive at the result:
\begin{equation}
\mathbf{\hat{\sigma}'} = \frac{\Lambda\hat{\sigma}\Lambda^T}{|(\Lambda^{-1})^T\hat{\mathbf{n}}| \det(\Lambda)}
\label{eq:My_Surface_Conductivity_Transform}
\end{equation}
where $\hat{\mathbf{n}}(r)$ is the local surface normal given by $\nabla F(\mathbf{r}) / |\nabla F(\mathbf{r})|$ and $\Lambda$ is the transformation matrix $\Lambda^{i'}_{i} = \partial{x^{i'}}/\partial{x^i}$.
The additional factor explicitly enters the surface conductivity tensor because a compression in the plane normal for the 2D material, should produce no transformation of the surface conductivity physically. But the $\det(\Lambda)$ factor does contain this compression factor. Therefore surface conductivity further requires a multiplicative factor for the renormalization of the surface delta function. Equivalently, one could say that the normal unit vector, after transformation, does not remain a unit vector. Hence the additional factor needs to be put in to ensure that in the transformed medium, $\mathbf{n'}$ is indeed a unit vector.

Taking cue from the SPP wave adapter proposed in \cite{doi:10.1021/nn200516r, Arigong:12}, we illustrate that the surface conductivity transformation indeed works, by using the transformation shown in Fig.\ref{fig:Schematic_Graphene_Transformation}. 

In the absence of any transformation optics, the plasmon mode propagating along the graphene sheet from the far left, would suffer substantial scattering into the free space modes. Such radiative loss is typically dependent on the radius of curvature of the bump. For instance, if the graphene plasmon mode is highly confined compared to the radius of curvature of the bump, then it is possible to achieve smaller radiative loss of the plasmon mode\cite{Lu:13}. In our current formalism however, we are able to tackle arbitrary radii of curvatures. To subvert this scattering one can employ the well known technique of transformation optics. This scheme would require us to modify the permittivity and permeability tensors in the region surrounding the sharp bump. However, just this would not be enough to prevent scattering since as we mentioned earlier, the surface conductivity also needs to be transformed in a way which depends on the details of the transformation we wish to carry out.

\begin{figure}
\begin{minipage}[l][10cm][t]{0.5\textwidth}
  \vspace*{\fill}
  \centering
  \includegraphics[width=8cm]{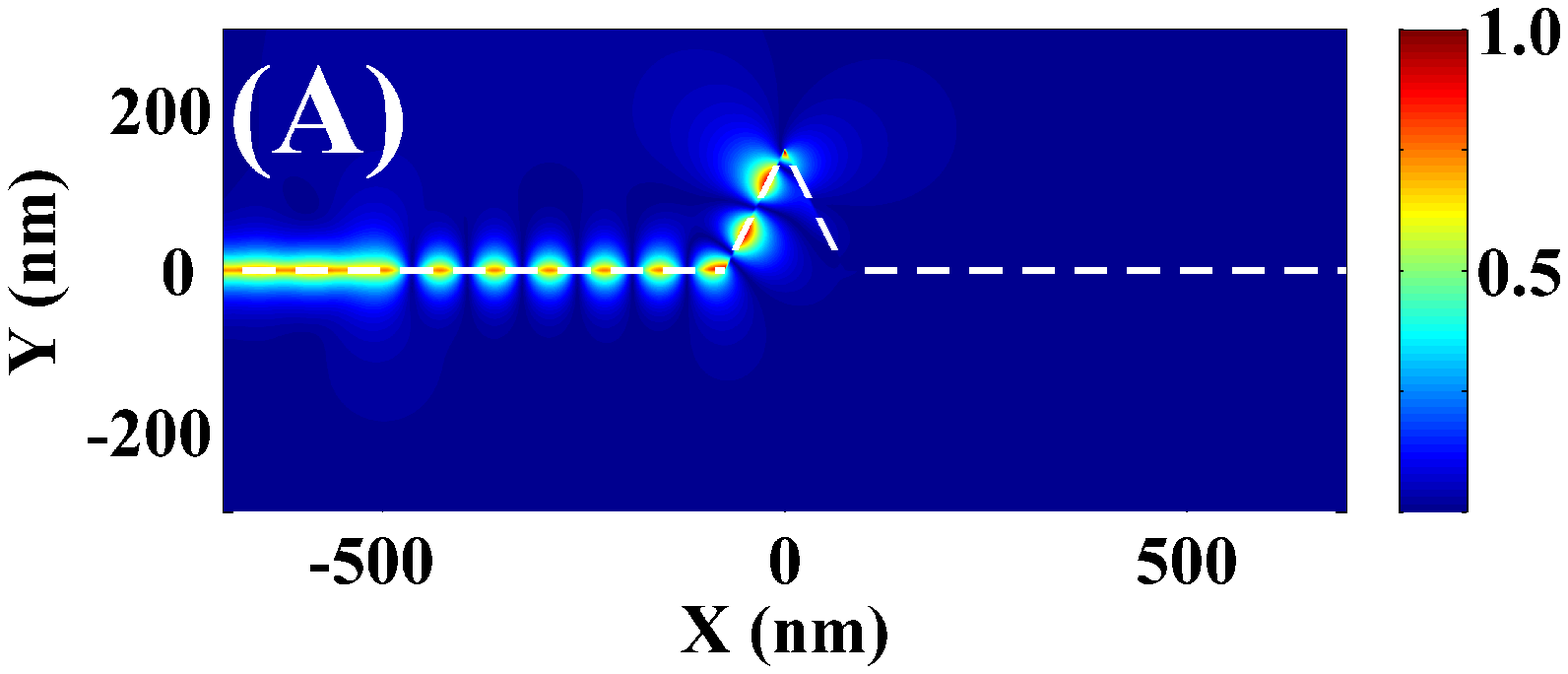}
  \label{fig:test1}\vfill
  \includegraphics[width=8cm]{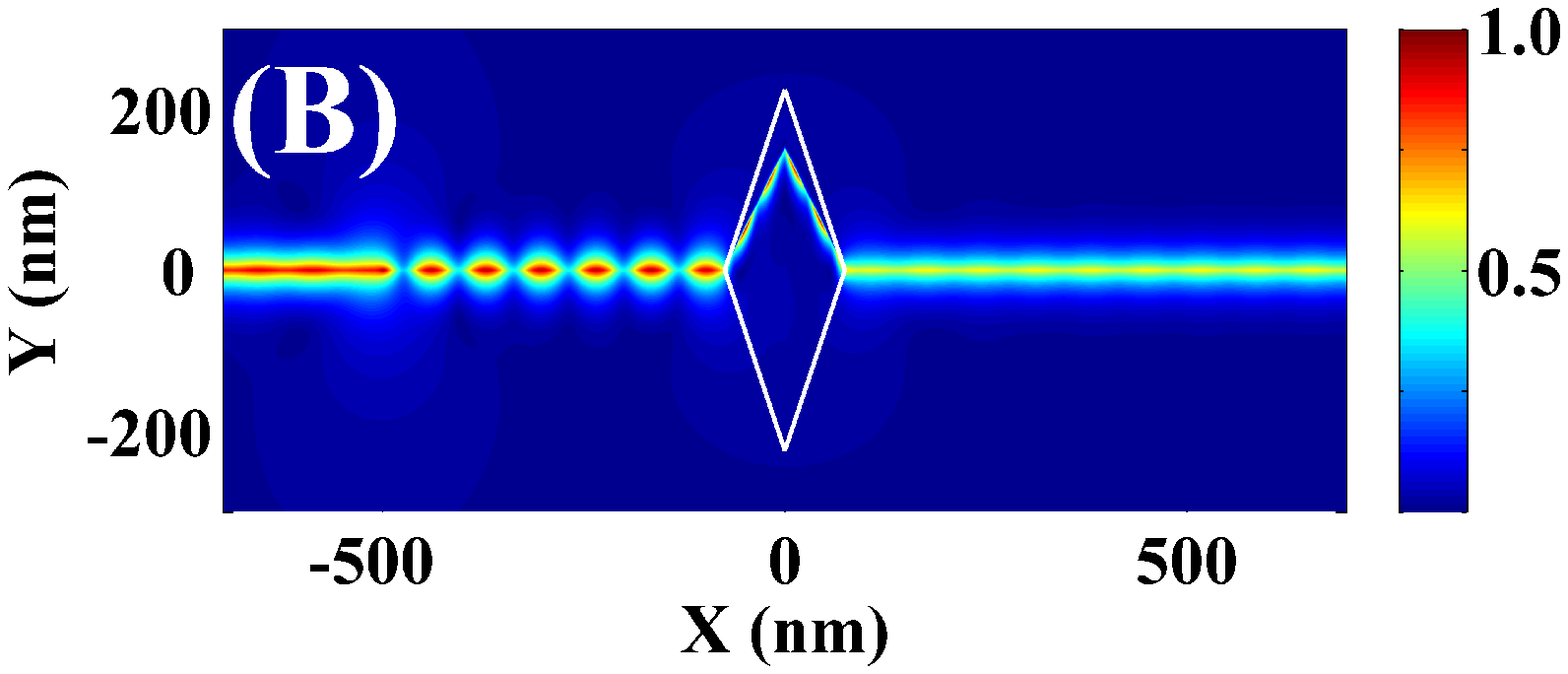}
  \label{fig:test2}\vfill
  \includegraphics[width=8cm]{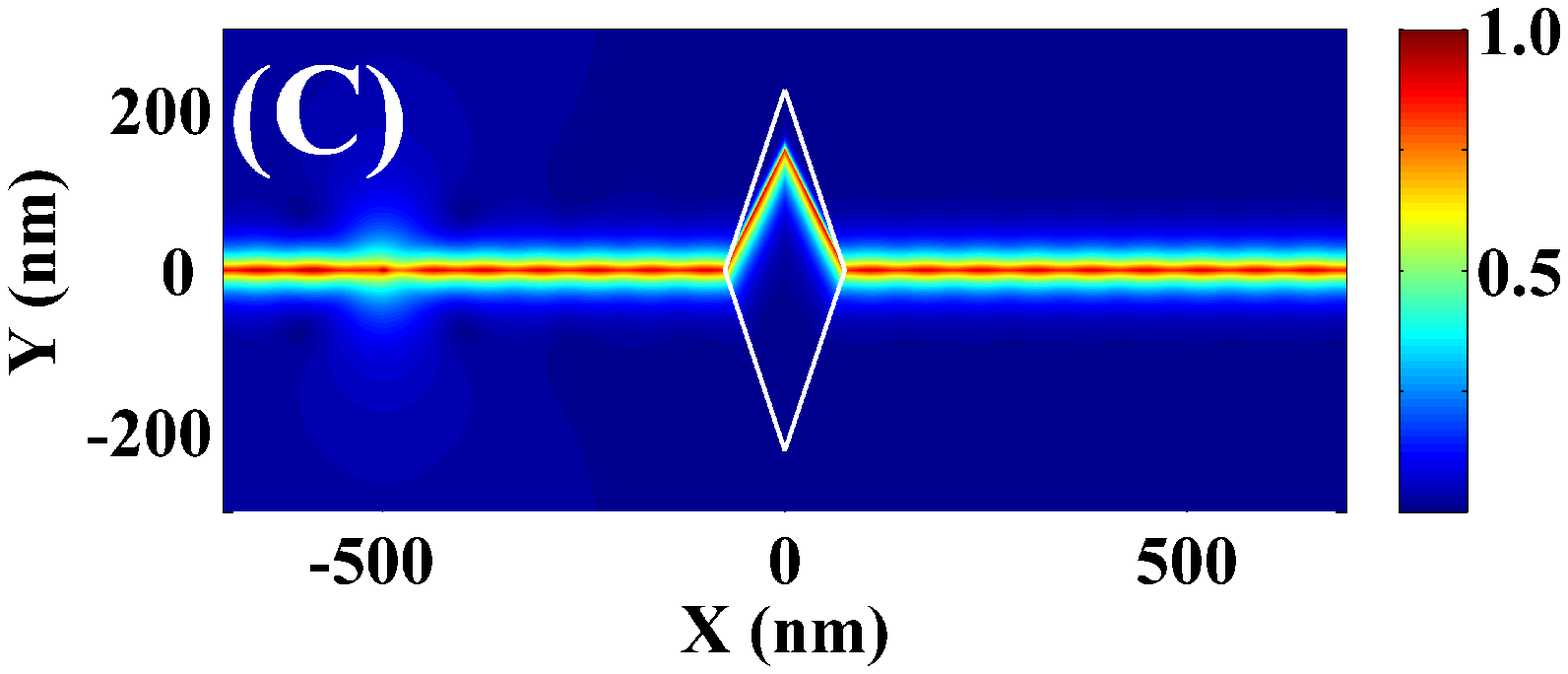}
  \label{fig:test3}\vfill
\end{minipage}
\caption{A magnetic line current source is placed at (-500nm, 2nm). Normalized scattered field $|\mathbf{H}-\mathbf{H_{inc}}|/\max\{|\mathbf{H}-\mathbf{H_{inc}}|\}$ is plotted here. A) Protrusion in graphene with no transformation optics, B) protrusion in graphene with transformation optics applied only to surrounding $\hat{\epsilon}$ and $\hat{\mu}$ and C) protrusion in graphene with the full transformation optics scheme applied to $\hat{\sigma}$, $\hat{\epsilon}$ and $\hat{\mu}$. In this example, $\alpha=\tan^{-1}3$ and $\beta=\tan^{-1}2$. In this example, the surrounding media in the untransformed case are assumed to be vacuum.}
\label{fig:FieldPlot}
\end{figure}

Hence we consider three cases: A) protrusion in graphene with no transformation optics employed, B) transformation optics employed but surface conductivity is not transformed and C) transformation optics employed for the surface conductivity as well as the bulk parameters. The results are shown in Fig.\ref{fig:FieldPlot}. Finite element simulations in Fig.\ref{fig:FieldPlot} were carried out using {\sc comsol multiphysics} by employing a surface current boundary condition to represent the graphene. The surrounding media in the un-transformed case is assumed to be vacuum, for this example. But we have performed tests, which are not presented here with substrates of non-unity refractive indices as well. For simplicity we use only the imaginary part of the surface conductivity of graphene at a doping level of $E_F = 0.5$ eV at zero temperature. The excitation frequency for this test case is $0.25$ eV, so the numerical value of the in-plane untransformed surface conductivity is $\sigma=\imath\ \SI{1.46e-4}{S}$.

Mathematically, the transformation in the four regions shown in Fig.\ref{fig:Schematic_Graphene_Transformation} is given as follows:
 \begin{flalign}
&x' = x \nonumber \\
&y' = y - \frac{\tan\beta}{\tan\alpha}|y| + (a-|x|)\tan\beta \\
&z' = z\nonumber
 \end{flalign}
Everywhere outside the four regions, there is no change. 
Note that we have chosen a linear map only for the purpose of demonstration of the main idea of surface conductivity transformation. The transformation we provided in Eq.\ref{eq:My_Surface_Conductivity_Transform} is valid in general, beyond the linear approximation. In principle, other transformations can be employed keeping in mind the material constraints. For instance, quasi-conformal maps\cite{PhysRevLett.101.203901}  which have been applied in the context of plasmonics elsewhere\cite{:/content/aip/journal/jap/114/14/10.1063/1.4824280}, could be employed;
or, in the case of rotationally symmetric three dimensional bumps of various shapes, it might be possible to design isotropic graphene-surface conductivity and the environment-permittivity profile obtained using techniques similar to \cite{PhysRevLett.111.213901}.

The Jacobian $\Lambda$ is given by:
\begin{equation}
\Lambda= 
 \begin{bmatrix}
 {1} & 0 & 0 \\
-\sgn(x)\tan\beta & 1-\sgn(y)\tan\beta/\tan\alpha & 0\\
0 & 0 & 1
\end{bmatrix}
\end{equation}
Then the transformed bulk material parameters in the four regions are given, in the $\{\Ket{x},\Ket{y},\Ket{z}\}$ basis, as follows:
\begin{widetext}
\begin{equation}
\epsilon', \mu'= \frac{\Lambda\Lambda^T}{\det \Lambda}=\frac{1}{1-\sgn(y)\tan\beta/\tan\alpha}
 \begin{bmatrix}
 {1} & -{\sgn(x)\tan\beta} & 0 \\
-{\sgn(x)\tan\beta} & {1-2 \sgn(y)\tan\beta/\tan\alpha + (\tan\beta/\sin\alpha)^2} & 0\\
0 & 0 & 1
\end{bmatrix}
\end{equation}
\end{widetext}
\begin{widetext}
The new surface conductivity tensor $\hat{\sigma}'$ is expressed using Eq.\ref{eq:My_Surface_Conductivity_Transform}, with $\hat{\mathbf{n}}=\hat{\mathbf{y}}$, as
\begin{equation}
\hat{\sigma}' = \frac{\Lambda\hat{\sigma}\Lambda^T}{|(\Lambda^{-1})^T\hat{\mathbf{y}}|\det \Lambda}=\sigma^{2D}
 \begin{bmatrix}
 \cos\beta & -{\sgn(x)\sin\beta} & 0 \\
-{\sgn(x)\sin\beta} & \sin\beta\tan\beta & 0\\
0 & 0 & \cos\beta
\end{bmatrix}
\label{eq:Specific_Transf_Surf_Cond}
\end{equation}
\end{widetext}
where $\sigma^{2D}$ is the scalar surface conductivity of the untransformed graphene.

Now, from Eq.\ref{eq:Specific_Transf_Surf_Cond}, it might appear that the surface conductivity tensor has off-diagonal components, which would be difficult to achieve experimentally. However, one must remember that the transformed tensor is given in the $\{\Ket{x}, \Ket{y}, \Ket{z}\}$ basis. To get more physical insight, we go the $\{\Ket{t}, \Ket{n}, \Ket{z}\}$ basis, where $\Ket{t}$ is a unit vector locally tangential to the graphene and $\Ket{n}$ is locally normal to graphene surface. Since the $z-$direction is unchanged in this example, we keep the same unit vector in that direction. A basis change can be carried out using $[\hat{\sigma}']_{tnz} = A[\hat{\sigma}']_{xyz}A^{\dagger}$, where $A$ is the basis change matrix composed of projections of the new basis vectors on the old basis vectors. In our case, graphene is situated in regions I and II, for which the $A$ matrix is given by
\begin{equation}
A = 
\begin{bmatrix}
 \cos\beta & -\sgn(x)\sin\beta & 0 \\
 \sgn(x)\sin\beta & \cos\beta & 0 \\
 0 & 0 & 1 
\end{bmatrix}
\end{equation}
Then $[\hat{\sigma}']_{tnz}$ is given by
\begin{equation}
[\hat{\sigma}']_{tnz} = \sigma^{2D}
\begin{bmatrix}
 \sec\beta & 0 & 0 \\
 0 & 0 & 0 \\
 0 & 0 & \cos\beta 
\end{bmatrix}
\label{eq:Transf_Surf_Cond_tnz_basis}
\end{equation}
Two things are apparent from the form of $[\hat{\sigma}']_{tnz}$ in Eq.\ref{eq:Transf_Surf_Cond_tnz_basis}. Firstly, there is a geometrical scaling factor involved in the transformed conductivity tensor. In particular, for the untransformed case, the components $[\hat{\sigma}]_{tt}$ and $[\hat{\sigma}]_{zz}$ should both be equal to $\sigma^{2D}$. These geometrical factors can be understood in terms of surface current conservation, as pointed out in \cite{PhysRevA.80.033820}. The surface current density in the tangential direction prior to the transformation is $J_{s, x} = \sigma^{2D}E_x$. After the transformation it becomes $J'_{s, t} = \sigma^{2D}\sec\beta(E'_x\cos\beta-\sgn(x)E'_y\sin\beta) = \sigma^{2D}\sec\beta (E_x\cos\beta)= J_{s, x}$. For the $z-$direction, we note that it is the surface current that needs to be preserved across the region $-a < x < a$. So we have, $I'_{s, z} = \int_{-a/\cos\beta}^{a/\cos\beta}J'_{s, z}dx'_t = \int_{-a}^{a}J'_{s, z}dx'/\cos\beta = \int_{-a}^{a}(\sigma^{2D}\cos\beta)E'_zdx'/\cos\beta = \int_{-a}^{a}\sigma^{2D}E_zdx = I_{s, z}$.

Secondly, it appears here that the given transformation requires us to somehow make the graphene conductivity anisotropic, that is, $\hat{\sigma}'_{tt}$ and $\hat{\sigma}'_{zz}$ need to be different. Although it might be possible to achieve such anisotropic effects using strained graphene\cite{0295-5075-92-6-67001, PhysRevB.81.035411, wright:163104, JMR:8477871}, yet in the present case of TM modes, $\hat{\sigma}'_{zz}$ component does not matter since the boundary condition for $H'_z$ only involves the surface current in the $\Ket{t}$ direction. Hence this conductivity change can be easily implemented via electrostatic\cite{:/content/aip/journal/apl/100/5/10.1063/1.3681799} or chemical doping. As far as the bulk parameters, namely $\epsilon', \mu'$ are concerned, there have been numerous demonstrations of natural and artificial anisotropic materials in the THz range\cite{bao:031910, bychanok:114304, ADMA:ADMA201103890, srep00078}.

We have demonstrated the transformation rule for surface conductivity under arbitrary coordinate transformations. An additional
factor related to the re-normalization of the surface current needs to be included to maintain the form invariance of the Maxwell's equations. We then presented an example problem of reducing scattering from a triangular protrusion in graphene, using the proposed method of surface conductivity transformation. This kind of conductivity transformation would be useful for transformation optics applications involving two dimensional materials.

 A.K. and N.X.F. acknowledge the financial support by the NSF (grant CMMI-1120724) and AFOSR MURI (Award No. FA9550-12-1-0488). K.H.F. acknowledges financial support from Hong Kong RGC grant 509813. We thank Professor Steven G. Johnson for helpful suggestions.

%\bibliographystyle{osajnl}
%\bibliography{OpEx_style}

%Create first reference list
\bibliographystyleref{osajnl_without_title}
\makeatletter
\renewcommand\@biblabel[1]{#1.}
\makeatother
\providecommand{\noopsort}[1]{}\providecommand{\singleletter}[1]{#1.}%

%begin the 5th page
\newpage

%start the new bibliography
%\nociteref{*}
\bibliographystyleref{osajnl}
\bibliographyref{OpexStyle}
\providecommand{\noopsort}[1]{}\providecommand{\singleletter}[1]{#1.}%

%\begin{thebibliography}{99}

%\end{thebibliography}

\end{document}